\documentclass[aps,pre,reprint,groupedaddress]{revtex4-1}

\usepackage{graphics}
\usepackage{graphicx}

\begin{document}


\title{Ice-lens formation and geometrical supercooling in soils and other colloidal materials.}


\author{Robert W. Style}
\email{style@maths.ox.ac.uk}

\author{Stephen S. L. Peppin}
\affiliation{Oxford Centre for Collaborative Applied Mathematics, University of Oxford, Oxford, UK}

\author{Alan C. F. Cocks}
\affiliation{Department of Engineering Science, University of Oxford, Oxford, UK}

\author{J. S. Wettlaufer}
\affiliation{Department of Geology and Geophysics, Department of Physics, \& Program in Applied Mathematics, Yale University, New Haven, Connecticut, USA}


\date{\today}

\begin{abstract}
 We present a new, physically-intuitive model of ice-lens formation and growth during the freezing of soils and other dense, particulate suspensions. Motivated by experimental evidence, we consider the growth of an ice-filled crack in a freezing soil. At low temperatures, ice in the crack exerts large pressures on the crack walls that will eventually cause the crack to split open. We show that the crack will then propagate across the soil to form a new lens. The process is controlled by two factors: the cohesion of the soil, and the geometrical supercooling of the water in the soil; a new concept introduced to measure the energy available to form a new ice lens. When the supercooling exceeds a critical amount (proportional to the cohesive strength of the soil) a new ice lens forms. This condition for ice-lens formation and growth does not appeal to any ad hoc, empirical assumptions, and explains how periodic ice lenses can form with or without the presence of a frozen fringe. The proposed mechanism is in good agreement with experiments, in particular explaining ice-lens pattern formation, and surges in heave rate associated with the growth of new lenses. Importantly for systems with no frozen fringe, ice-lens formation and frost heave can be predicted given only the unfrozen properties of the soil. We use our theory to estimate ice-lens growth temperatures obtaining quantitative agreement with the limited experimental data that is currently available. Finally we suggest experiments that might be performed in order to verify this theory in more detail. The theory is generalizable to complex natural-soil scenarios, and should therefore be useful in the prediction of macroscopic frost heave rates.
\end{abstract}

\pacs{46.05.+b,46.50.+a,82.70.Dd,83.80.Hj}

\maketitle

\section{Introduction}

Frost heave is an incredibly powerful process, capable of molding landscapes anywhere that temperatures drop below $0^\circ$C. During freezing, changes in soil volume cause the surface to swell and this can result in a variety of fascinating patterns being sculpted into periglacial landscapes \cite{will91,dash06}. These take many forms, such as stone circles, polygons and labyrinthine shapes \cite{kess03}, and occur on Earth as well as on the frozen surface of Mars \cite{thom01}. Frost heave also leaves its mark after frozen soils have melted. Damaged rocks can break away to kickstart landslides, while topsoils can be washed downslope by meltwater. Even flat, frozen soils often collapse unevenly when melting, resulting in hummocked ground and distinctive `drunken forests' where trees fall, or are left at unnatural angles due to soil movement. This `thaw weakening' is also well-known as a major cause of structural damage to roads and buildings in cold climates and has a significant cost, in particular because of the regular requirement to repair roads after freezing. For instance, in 1999 it was estimated that in the USA over two billion dollars was spent on reparations to roads caused by frost-heave damage alone \cite{dimi99}.

In order to model the macroscopic effects of frost heave it is important to understand the micro- and meso-scale physics that underlie this process. However, despite extensive prior research, there are still important questions remaining about how frost heave occurs. It is known that frost heave is not due to the expansion of water upon freezing (as commonly assumed); frost heave still occurs when water is replaced by substances that contract upon solidification \cite{tabe29}. Instead heave occurs when segregated ice forms within a column of soil that has its base in an unfrozen liquid reservoir. Thermomolecular pressure gradients then cause liquid to be drawn up from the relatively warm reservoir towards the cold, segregated ice whereupon it freezes (this process is commonly referred to as `cryosuction') \cite{wors99}. Thus the soil `expansion' is due to the added mass of water brought into the soil column from the reservoir (e.g. \cite{remp04}).

In the last forty years there has been much experimental work aiming to elucidate the mechanisms at work in the ice segregation process and it is now known that ice can segregate in a range of morphologies \cite{pepp07b,pepp08}. However the most well-known configuration is the periodic ice-lens formations that were first observed in the early experiments of Taber \cite{tabe29,tabe30}.
It is the growth of these ice lenses that determines the soil's heave rate. Therefore it is important for us to understand how they form in order to make accurate, quantitative predictions of the frost-heave process. Previous theories of ice-lens formation have been proposed by O'Neill \& Miller \cite{onei85}, Gilpin \cite{gilp80} and Rempel et al. \cite{remp04} which have provided the foundation for most subsequent work on frost heave (e.g. \cite{fowl89,chri06}). However the theory of O'Neill \& Miller is effectively empirical, as it is based on a postulated `stress partition function' that is physically-unjustifiable. Rempel et al. \cite{remp04} provided a microphysically-correct model for the growth of periodic ice lenses. They, and Gilpin consider how periodic ice lenses can form in the presence of a `frozen fringe' (a region, ahead of the warmest ice lens, where liquid freezes into ice in the micron-sized pores of the soil). However in some systems, such as colloidal clays, it is known that periodic ice lenses can form when no frozen fringe is present (e.g. \cite{wata00},\cite{brow90}). In order to explain these observations, we present here a new, physically-based theory that shows how multiple lenses can form, even in the absence of a frozen fringe. Moreover, we show that this model is in good qualitative and quantitative agreement with experiments on ice-lens formation in a variety of colloidal materials. Throughout the paper, we shall refer to the material as a soil, however we note that this theory is equally applicable to general colloidal materials.

\section{A new model for ice-lens formation and growth}

\begin{figure}
\centering
\includegraphics[width=5cm]{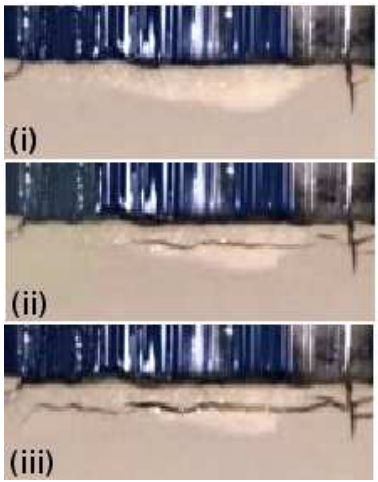}
\caption{(Color online) Ice lens formation in freezing kaolinite clay. The clay is being frozen from the top downwards in the apparatus used by Peppin et al. \cite{pepp07}. (i)-(iii) are photographs taken 20 minutes apart in time-lapse footage of ice-lens growth. The dark, upper region of each photograph is the previous ice lens. The light, lower region is kaolinite clay. A new ice lens (dark, crack-like) can be seen propagating from right to left through the clay. A vertical shrinkage crack can also be seen on the right-hand side of the photographs.}
\label{fig:wata}
\end{figure}

In order to formulate a new model of ice-lens growth, we draw upon intuition gained from a series of experimental observations which indicate that the initial segregation of ice in freezing materials is seemingly `crack-like' in nature. Firstly, Chamberlain \& Gow \cite{cham79} and Arenson et al. \cite{aren08} froze water-saturated soils and, by taking cross-sections of the frozen material, observed the presence of ice-filled shrinkage cracks in a polygonal pattern. They noted that these cracks are caused by the development of very low pressures (due to cryosuction) in much the same way that desiccation cracks form on the surface of rapidly-drying clays. Secondly, recent experiments have enabled us to closely observe growing ice lenses in order to investigate the nucleation process of subsequent lenses. Figure \ref{fig:wata} shows three snapshots taken during the propagation of a new ice lens across a section of a freezing cell in our experiments on the freezing of kaolinite clay. The cell is cooled from above. The upper half of each picture shows the last (warmest) ice lens, while the lower (light-colored) half is kaolinite clay. A new lens can be seen growing across the cell from right to left in a remarkably crack-like manner (the new lens grows approximately 1mm in front of the old lens, and the frames are taken 20 minutes apart). Similar observations have been made in experiments on the freezing of silica microspheres by Watanabe et al. \cite{wata00} (Watanabe, private communication). Thirdly, Akagawa et al. \cite{akag06} performed multiple frost-heave experiments on a sample of Dotan silt. They concluded that the tensile strength of the soil strongly affects ice-lens formation. Ice-lens formation pushes soil particles apart, so that when the frozen soil is melted, there is a lower tensile strength at the points where previous lenses reformed. In subsequent freezing experiments, ice lenses formed at these weakened points. Interestingly, despite forming ice lenses at the same points, less undercooling was required for new ice lenses to form in later experiments. Akagawa et al. \cite{akag06} suggest that this is due to the fact that the forces acting to open an ice lens must overcome the tensile strength of a soil before an ice lens grows. Thus when the tensile strength is lower, it is easier to form ice lenses (we also note that ice-lens nucleation will be promoted by increased soil porosity that may exist at the position of previous lenses). Further evidence supporting this interpretation was obtained by performing freezing tests on the same material, broken up and reconsolidated in order to reduce the tensile strength of the soil. In this case inter-lens spacing was reduced by a factor of 5, suggesting that ice lenses are significantly easier to nucleate when the soil tensile strength is reduced.

These experiments indicate the strength of the soil is an important factor, and that ice-lens nucleation may be caused by material fracture in the presence of internal stresses. Thus we take a fracture-mechanics approach to the problem. We consider the stresses on a small flaw in the material, and calculate at what point these stresses become sufficiently large to overcome interparticle cohesion and cause the flaw to enlarge. As we shall see, this will allow us to obtain a new, physically-intuitive condition for ice-lens nucleation. We should note that fracture mechanics has been previously used by Walder and Hallet \cite{wald85} to model segregated-ice growth in freezing rocks, a process which has been recognized as being related to ice-lens formation in soils. However, we will take a different approach to that work, in particular by not assuming the presence of a frozen fringe.

\begin{figure}
\centering
\includegraphics[width=8cm]{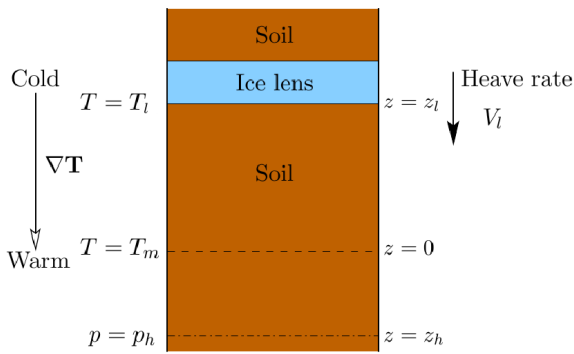}
\caption{(Color online) Schematic diagram for the freezing of clay.}
\label{fig:schem}
\end{figure}

In order to clearly demonstrate the essential physics underlying our theory of ice-lens formation, we consider a simple, one-dimensional model of frost heave as shown in figure \ref{fig:schem}. It will be seen later that the model can be straightforwardly generalized to more complex settings. For continuity and transparency we base our notation on that of Rempel et al. \cite{remp04}. The model is based upon laboratory frost-heave experiments using directional-solidification cells such as have been performed by several groups (e.g. \cite{pepp08,wata00}). A saturated soil or a colloidal suspension is placed in a cell in which a fixed temperature gradient $\nabla T$ is imposed. This `frozen temperature' approximation is commonly invoked in modeling direction-solidification experiments, and is based on having a cell with a large thermal mass in comparison to that of the material contained within it. Then temperature fluctuations caused by latent heat release on freezing, and due to the interphase variation in thermal conductivity have little impact on the overall temperature gradient in the system (e.g. \cite{davi01}).

We define the vertical co-ordinate $z$ so that $z=0$ when $T=T_m$, the bulk melting temperature of pure water. The imposed temperature is below freezing for $z>0$, and hence at some point above $z=0$ the soil will begin to freeze. In a typical experiment a sequence of ice lenses form across the cell. The position of the bottom of the warmest ice lens is given by $z_l$, and the temperature at this point is $T_l$.
We do {\emph{not}} assume the existence of a frozen fringe and hence on the warm side of the warmest ice lens is a region of soil that may or may not be partially frozen. The reason that the soil can remain unfrozen below the melting point is due to the Gibbs-Thomson effect whereby the melting point of an ice surface convex into its melt is reduced relative to that of a planar ice-water interface \cite{dash06}. This prevents ice from invading small radius pores until a certain undercooling of the soil is reached \cite{remp04}. Regardless of the existence of a frozen fringe, provided that there is an interconnected liquid network in the porous medium, we can measure the Darcy pressure $p$ (see Appendix A for detailed discussion). At some depth $z_h<0$, we assume that there is a warm liquid reservoir at which the Darcy pressure can be assumed to be a constant, $p_h$.
In addition there is an overburden pressure $P_o$ exerted on the upper surface of the soil (in field situations this corresponds to the weight of material overlying the soil) and so mechanical equilibrium tells us that the ice pressure within any lens that extends across the cell is given by $P_o$.

A useful relationship between the Darcy pressure of pore water $p$ and the pressure $P$ of ice in local equilibrium with the water is given by the Clapeyron equation
\begin{equation}
P-p=\rho L_m \frac{(T_m-T)}{T_m}.
\label{eqn:clap}
\end{equation}
Here $\rho$ is the density of the solid/liquid phase (we ignore density differences between ice and water) and $L_m$ is the latent heat of melting of ice \cite{blac95}. Care needs to be taken in applying this equation to ice in freezing porous media because strictly-speaking the equation is derived for liquid in equilibrium with {\em bulk} ice. In effect this means that it should be applied to ice that takes up a volume much larger than a particle size such as that in an ice lens or in a large flaw (see Appendix A for a discussion). Thus care must be taken in examining ice inside typical soil pores that are around the same size as a soil particle. As an example of the use of this equation we know that the pressure in an ice lens is $P_o$, as it is in mechanical equilibrium with the overlying weight given by the overburden pressure. Therefore assuming that the liquid adjacent to an ice lens is in local bulk equilibrium with the ice lens, we find that the Darcy pressure of liquid at an ice lens is given by $p=P_o-\rho L_m(T_m-T)/T_m$. We note that strictly the Clapeyron equation only holds when the growth rate of the ice is slow, and kinetic effects need to be added at higher growth rates. The implications of these effects for this model will be addressed in a forthcoming work.

Here, we assume the existence of `large flaws' within the soil in order to be able to use the Clapeyron equation (\ref{eqn:clap}). By `large' we shall take it that the length of the flaw is as least ten times larger than the radius of the constituent particles of the soil. Such flaws are commonly assumed to exist in the analysis of material fractures \cite{ande05}, and they have been observed experimentally in a variety of clays. For instance Diamond \cite{diam70} observed pore sizes of larger than 100$\mu$m in kaolinite, while scanning electron micrographs showed individual clay particles to be $O(1\mu$m$)$ in size. Similar results were obtained for bentonite, illite, and a range of silts.

\begin{figure}
\centering
\includegraphics[width=8cm]{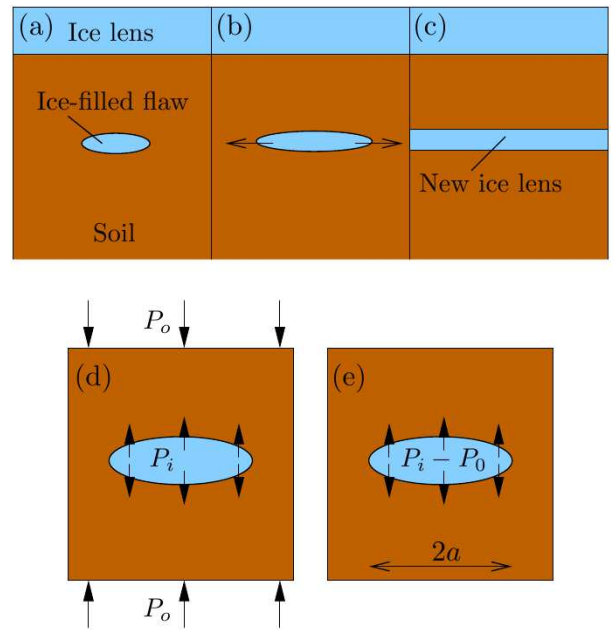}
\caption{(Color online) The opening of a flaw in a soil. (a-c) The process by which a large, ice-filled flaw in the soil ahead of a frozen fringe is pushed open to form a new ice lens. (d) Pressures acting on the system around the flaw. (e) Pressures acting on the flaw after linear superposition to remove the farfield stress.}
\label{fig:flaw}
\end{figure}

Within this framework let us consider the conditions under which a new ice lens will form. Consider a large flaw ahead of the warmest ice lens and at a temperature below $T_m$, as shown in figure \ref{fig:flaw}(a). We assume that ice is present within the flaw (we will return to the question of how this occurs in section \ref{section:51}). As mentioned above, the flaw is assumed to be large compared to a soil particle of size $R$ so that we can make use of the Clapeyron equation (\ref{eqn:clap}), while we also assume that it is small relative to the distance to the nearby ice lens and to the size of the fully grown lens. The Clapeyron equation shows that, because the undercooling $T_m-T$ is positive, the pressure of the ice within the flaw, $P_i$, will be larger than the Darcy pressure $p$. Therefore as the undercooling becomes large, $P_i$ will become large and positive, and so the ice within the flaw will exert a pressure on the walls of the flaw. Eventually, this pressure will overcome the cohesive strength of the soil, and the flaw will extend. Later, we shall see that the flaw will grow parallel to the previous ice lens. Thus, as the flaw extends, it cuts across the width of the sample with the result being the formation of a new ice lens. This is shown schematically in figure \ref{fig:flaw}(a-c), with the vertical extent of the flaws being overexaggerated for clarity (in practice the vertical opening will be much smaller than the length of the crack). We note that there will actually be many such flaws, each at different orientations to the previous ice lens. However (as can be derived from the analysis of stress that follows later in this paper) the crack-opening stresses will be greatest on the flaws that are orientated parallel to the previous lens, and it is these flaws that will be the first to extend to form new ice lenses. Thus we only need consider flaws of this particular, `horizontal' configuration. Also, note that we use a single, isotropic pressure to characterize the ice stresses in the flaw, rather than concerning ourselves with the possibility of differential stresses arising in the ice. In fact, as explained in Appendix A, because the ice nucleates from the liquid (which can maintain no differential stresses) we expect the ice in a flaw to remain under isotropic stress so that it can be characterized by the pressure $P_i$.

We assume that the material around the flaw can be treated as a linear elastic solid. This has the benefit of allowing us to use linear-elastic fracture mechanics to investigate the conditions under which the crack will extend. Unfrozen, overconsolidated soils of high clay concentrations have been observed to have substantial linear-elastic responses (e.g. \cite{grah83,muir90}) and so this is a reasonable first approximation to the material behavior. When there is pore ice in the soil (i.e. in a frozen fringe) its presence will further reduce viscous and plastic behavior of the soil and so the linear-elastic approximation is again expected to be reasonable. In general, the elastic moduli of the system will change with $z$ due to changes in particle packing fraction and ice fraction. However initial flaws will be much smaller than the lengthscale over which these moduli are expected to change. So in conducting an analysis of the stresses acting on the flaw, we can take the flaw as being surrounded by a homogeneous elastic material.

As discussed above, the pressure $P_i$ of the bulk ice in the flaw is related to the local Darcy pressure $p$ by the Clapeyron equation (\ref{eqn:clap}). Therefore if we know the local temperature and Darcy pressure at the crack position, its internal pressure is $P_i=p+\rho L_m (T_m-T)/T_m$. In addition the far-field stress acting on the soil is given by the overburden pressure $P_o$. This means that the stresses acting on the system reduce to those shown in figure \ref{fig:flaw}(d): an internal pressure acting to open the crack $P_i$, and the overburden pressure acting to close the crack. We wish to ascertain whether this system of stresses can drive crack extension and thus ice-lens formation. Because the material is assumed to be linear-elastic, we can use linear superposition to show that this problem is equivalent to the case of a flaw under internal pressure $P_i-P_o$, as shown in figure \ref{fig:flaw}(e) \cite{ande05}.

According to linear-elastic fracture mechanics the flaw will grow when the stress intensity factor $K_I$ (the magnitude of the stress singularity at the flaw tip) exceeds the critical value $K_{Ic}$, a material property measuring the fracture toughness of the soil. In the case shown in figure \ref{fig:flaw}(e) the stress intensity factor is given by
\begin{equation}
K_I=(P_i-P_o)\sqrt{\pi a},
\label{eqn:sif}
\end{equation}
where $a$ is half the length of the crack (e.g. \cite{ande05}). Therefore the flaw will extend when
\begin{equation}
P_i-P_o=\frac{K_{Ic}}{\sqrt{\pi\lambda R}}\equiv \sigma_t.
\label{eqn:stresscond}
\end{equation}
Here we have set $a=\lambda R$, where $\lambda \gtrsim 10$ so that the size of the flaw is much bigger than the particle size. We also recognize that the right-hand side of this equation can be interpreted as the tensile strength of the soil $\sigma_t$. Note that as the crack grows, the stress intensity factor increases. This means that when the internal pressure is sufficiently large to cause a crack to grow, the crack growth will not halt. Instead as the crack grows, new ice will freeze in the newly-created crack volume, pressurizing the walls of the crack with the ice pressure $P_i$. The stress intensity factor will then exceed $K_{Ic}$ (as the new crack is longer than the original crack), giving rise to further extension. This positive feedback means that the crack will continue to grow through the material until it reaches the material boundary (catastrophic failure). At that point, the ice-filled crack becomes the new warmest ice lens (figure \ref{fig:flaw}(c)). Additionally we may also note that the fact that larger cracks have larger stress intensity factors means that the largest cracks will break open to form ice lenses first. This means that in calculating the conditions under which a new ice lens will form, we only need to consider the growth of the largest flaws present in the system, and need not worry about smaller pores with $\lambda<10$.

As with other fracturing events, crack growth is expected to be rapid in comparison to the vertical growth rate of an ice lens. As an example, in our directional-solidification experiments on kaolinite clay, time-lapse footage suggests that the fracture extension velocity is at least a factor of 10 times faster than the ice growth rate. This factor is likely to be substantially larger but is currently constrained by the time between frames in the available experimental data.  Therefore it is reasonable to pinpoint the position of the new ice lens to the point at which the stress intensity factor first becomes large enough that the crack starts to extend. This means that equation (\ref{eqn:stresscond}) will represent the point at which a new ice lens will form. This new condition can be compared to the previously-assumed condition that ice lenses will first grow when the interparticle pressure between soil particles first becomes zero or reaches a critical negative value (e.g. \cite{onei85,fowl89,remp04}) and it will be seen that these are not equivalent. However, it is interesting to note that Akagawa et al. \cite{akag06} have previously suggested that a condition of the same form as equation (\ref{eqn:stresscond}) gives good agreement with their experiments on ice-lens nucleation.

\section{The criterion for ice-lens nucleation \label{disc}}

\subsection{Geometrical supercooling}

Consider a parcel of liquid at temperature $T$ and Darcy pressure $p$ ahead of (warmer than) the warmest ice lens. Suppose a new ice lens were to be inserted next to the liquid such that the liquid and the ice are in local equilibrium. The pressure of the ice lens will be $P_o$, and if the ice and liquid are to be in equilibrium they must be at the temperature $T_{\ell}$ (the {\em ice-lens growth temperature}) that from equation (\ref{eqn:clap}) is given by
\begin{equation}
 T_{\ell}= T_m\left(1-\frac{P_o-p}{\rho L_m}\right).
\label{eqn:Til}
\end{equation}
We can compare the actual temperature of the parcel of liquid $T$ to the ice-lens growth temperature $T_{\ell}$. If the temperature is below the ice-lens growth temperature, then energy can be released if a new ice lens forms. Thus, in principle, in a soil with no cohesive strength a new ice lens can form wherever $T\le T_{\ell}$. However, in cohesive soils ice lenses will not form spontaneously as soon as $T=T_{\ell}$ because formation of an ice lens is hindered by the difficulty of breaking apart the soil structure to accommodate the new lens. Hence it is possible for the temperature of the soil to be less than $T_{\ell}$ without a new ice lens forming; this can be thought of as being supercooled relative to the ice-lens state. We shall call this {\em geometrical} supercooling. We show in Appendix B that this is analogous to constitutional supercooling that occurs in the freezing of alloys and dilute particle suspensions.

\subsection{Nucleation criterion}

Now we return to the ice-lens nucleation criterion (\ref{eqn:stresscond}). Substituting equations (\ref{eqn:clap}) (with $P$ given by the pressure of ice in the flaw $P_i$) and (\ref{eqn:Til}) into (\ref{eqn:stresscond}) we obtain
\begin{equation}
P_i-P_o=\frac{\rho L_m(T_{\ell}-T)}{T_m}=\frac{K_{Ic}}{\sqrt{\pi\lambda R}},
\label{eqn:pipo}
\end{equation}
and with some rearrangement we find that the condition for ice-lens initiation is
\begin{equation}
T_{\ell}-T=\frac{T_m K_{Ic}}{\rho L_m \sqrt{\pi \lambda R}}=\frac{T_m\sigma_t}{\rho L_m}.
\label{eqn:supercool}
\end{equation}
This reinforces the argument, outlined above, that in order for the soil to fracture and nucleate a new ice lens, there must be geometrical supercooling of the pore liquid. When this supercooling reaches a critical amount, given by the right hand side of equation (\ref{eqn:supercool}) the soil breaks apart and a new ice lens is formed. Note that this critical value is likely to depend on the initial packing of the soil particles (e.g. \cite{akag06}), as this will alter the flaw sizes present in the material. As a check of intuition, it can be seen that when there is no cohesion of the soil around the crack, $K_{Ic}=0$ and a new lens will form at the point when the pore liquid becomes supercooled. In fact, we shall see later that if $T_{\ell}-T<0$ and a new ice lens forms, the ice lens would melt, so in the absence of geometrical supercooling we would expect to see a single ice lens or needle ice at the freezing surface of a soil (primary heaving) \cite{wors99}. This suggests that the presence of geometrical supercooling is intimately linked to the appearance of new ice lenses in a freezing soil.

\begin{figure}
\centering
\includegraphics[width=6cm]{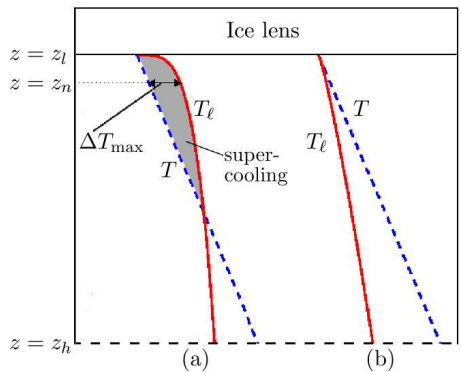}
\caption{(Color online) Schematic diagram of the soil conditions that can occur under a growing ice lens. The continuous curves correspond to the ice-lens growth temperature, while dashed curves represent the actual temperature of the soil. (a) Geometrical supercooling occurs. Ice lens nucleation can occur at the first moment when the geometrical supercooling ($\Delta T_{\rm{max}}$) is sufficiently large for the fracture condition (\ref{eqn:supercool}) to be satisfied. A new lens will form at this point ($z_n$). This type of profile for $T_{\ell}$ will typically arise during the diffusive evolution of the pore pressure field after rapid cooling at the ice lens, analogously to the appearance of constitutional supercooling in alloys \cite{wors86}. (b) No geometrical supercooling occurs below the ice lens and no new ice lenses can form. This will typically occur at slower rates of cooling.}
\label{fig:supercool}
\end{figure}

Figure \ref{fig:supercool} shows the two representative situations that can potentially occur in a freezing soil. The figures show profiles of the temperature $T$ and the freezing temperature $T_{\ell}$ in the soil. For simplicity we assume that the temperature is a linear function of $z$, though this will not be true in general. At the bottom surface of the warmest ice lens ($z=z_l$), the bulk ice of the lens is in local thermodynamic equilibrium with the pore water and so $T=T_{\ell}$. At the water reservoir position ($z=z_h$) the water is warmer than the bulk melting temperature $T_m$ and so $T>T_{\ell}$. This can be seen from equation (\ref{eqn:Til}) along with assuming that the overburden pressure is greater than or equal to the reservoir pressure $p_h$, as will generally be the case in typical frost heave situations. Figure \ref{fig:supercool}(a) shows the case where $|d T/d z|>|d T_{\ell}/d z|$ and so geometrical supercooling occurs whereas figure \ref{fig:supercool}(b) shows the case where $|d T/d z|\leq|d T_{\ell}/d z|$ and so no geometrical supercooling occurs. 

In the case where there is geometrical supercooling there will be a point at which the supercooling reaches a maximum value, $T_{\ell}-T=\Delta T_{\rm{max}}$, as shown in \ref{fig:supercool}(a). In general, $\Delta T_{\rm{max}}$ will increase as the freezing progresses. Therefore after some time $\Delta T_{\rm{max}}$ will exceed the critical supercooling $T_m K_{Ic}/\rho L_m \sqrt{\pi \lambda R}$ and a new ice lens will grow at the point of maximum supercooling, $z=z_n$. This occurs a finite distance ahead of the previous lens, and so this provides a mechanism by which periodic, discreet ice-lens patterns can occur.

New cracks propagate parallel to previous ice lenses because the internal pressure driving the crack growth is greatest at $z=z_n$. Thus the greatest amount of energy is released for cracks that continue to propagate at this height. In addition, such horizontal cracks have no shearing component, which means that the crack should propagate in a straight line without kinking out of plane \cite{ande05}. This agrees with the general behavior seen in typical frost-heave experiments. Note that we only consider here an ideal one-dimensional system with no features such as shrinkage cracks or boundaries in the horizontal direction. It is likely that in the field and the laboratory, there will be more complex stress conditions in the soil due to soil imperfections, and interaction between the ice lenses and other features. We assume that these additional stresses will then help to explain the variations in ice-lens configurations that are sometimes observed experimentally (e.g. \cite{cham79}). 

Qualitative conditions under which ice lenses form can be determined from the fact that for supercooling to occur $|d T_{\ell}/d z|$ must be large near the ice lens, and drop off as the soil becomes warmer (see figure \ref{fig:supercool}). From equation (\ref{eqn:Til}), $P_o-p=\rho L_m (T_m-T)/T_m$ and so
\begin{equation}
\frac{d T_{\ell}}{d z}=\frac{T_m}{\rho L_m}\frac{d p}{d z},
\label{eqn:tp}
\end{equation}
while the flow of liquid in the soil is controlled by Darcy's law
\begin{equation}
U=-\frac{k}{\mu}\frac{d p}{d z},
\label{eqn:darcy}
\end{equation}
where $U$ is the total volume flux of water through the soil, $k$ is the permeability of the soil and $\mu$ is the dynamic viscosity of water. Combining equations (\ref{eqn:tp}) and (\ref{eqn:darcy}) gives
\begin{equation}
\frac{d T_{\ell}}{d z}=-\frac{\mu U T_m}{\rho k L_m}.
\label{eqn:dTil}
\end{equation}
Considering the case where $U$ is approximately constant throughout the depth of the soil column, it can then be seen that the only term on the right hand side of this equation that is not a constant is the soil permeability $k$. So when the permeability is small, $|dT_{\ell}/dz|$ is large and vice versa. Therefore ice-lensing conditions are expected to be found when $k$ is small directly ahead of the warmest ice lens, and increases towards the water reservoir at $z=z_h$. There are three major situations in which this can occur. Firstly, if ice enters the pore space directly below the ice lens to create a frozen fringe, the permeability will be drastically decreased there due to the reduction in pore volume of water (c.f. Hansen-Goos \& Wettlaufer \cite{hans10}, whose work also shows how even small amounts of impurities will play a large role in determining the permeability after the formation of a frozen fringe). In particular, frozen fringes are expected to occur in soils with larger pore sizes. Secondly, if the soil matrix is very compressible, particle rejection in front of the ice lens compresses the particles together leading to a reduction in porosity and permeability (e.g. \cite{pepp06}). This is expected to occur in swelling clays like bentonite, which has been observed to remain unfrozen down to $-8^\circ$C \cite{brow90}. Thirdly, desaturation of the soil adjacent to the growing ice lens can occur due to the rejection of dissolved air at the growing ice lens and to the large suctions that can develop within the soil because of the low temperature (e.g. \cite{brow90,or02}). Desaturation was identified as an important factor for consideration in the frost-heaving system by Penner \cite{penn63}, and its occurrence can lead to reductions in permeability by several orders of magnitude \cite{fred93}. We have observed evidence of desaturation in our own experiments on the freezing of kaolinite clay, and will report on this in future work. Any of these three scenarios provide a viable mechanism for geometrical supercooling to occur, and thus the occurrence of frost heave.

\section{Heave rate surging}

\begin{figure}
\centering
\includegraphics[width=8cm]{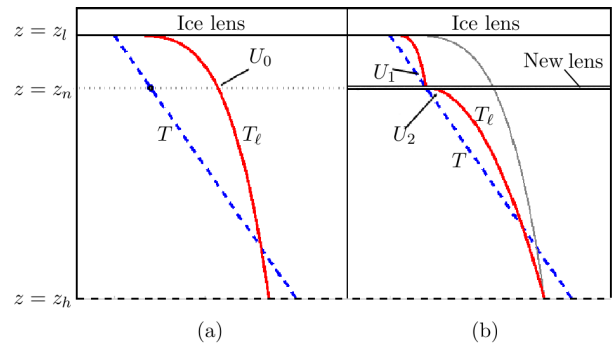}
\caption{(Color online) Schematic diagram showing the changes occurring in the soil upon ice nucleation. The dark, continuous curves represent the freezing temperature of the pore water, dashed curves represent the actual temperature of the soil. (a) Before an ice lens forms, $\Delta T_{\rm{max}}$ is sufficiently large that ice lens nucleation will occur at the point $z_n$. $U_0$ is the volume flow rate of liquid at $z_n$. (b) After ice lens growth $T_{\ell}$ becomes equal to $T$ at the ice lens. This alters the flow rate at $z_n$ so there is a discontinuity between $U_1$ and $U_2$, the volume flow rates of liquid above and below the new ice lens respectively. As discussed in the text, the volume flow rates are proportional to the gradient of $T_{\ell}$. The light, continuous curve corresponds to $T_{\ell}$ before lens formation occurs.}
\label{fig:nucleate}
\end{figure}

It is worth devoting some consideration to what occurs when a new ice lens nucleates, in order to make predictions of the change in heave rate during the period of formation of a new ice lens. Figures \ref{fig:nucleate}(a) and (b) show the situation immediately before the growth of a lens, and immediately after the lens has grown across the width of the system. As soon as the ice lens grows, the geometrical supercooling vanishes at the position of the new ice lens so that the pore water is in local thermodynamic equilibrium with the ice in the new lens. This means that $T_{\ell}=T$ at the new lens and the $T_{\ell}(z)$ profile can be recalculated using equation (\ref{eqn:dTil}), as we have shown qualitatively in figure \ref{fig:nucleate}(b). From equation (\ref{eqn:Til}) $T_{\ell}$ is linearly dependent on the Darcy pressure $p$, so the drop in $T_{\ell}$ at the position of the new ice lens corresponds to a drop in the liquid pressure there. This means that the adjustment of the $T_{\ell}$ profile can be thought of as the adjustment of the soil liquid to the new (lower) Darcy pressure at the ice lens. 

Equation (\ref{eqn:dTil}) shows how the volume flux of liquid in the soil is related to the gradient of the liquid freezing temperature. Before the ice lens nucleates, the volume flux at the position of the new ice lens $U_0\equiv U(z_n)$ is continuous because $dT_{\ell}/dz$ is continuous (we assume $k$ is a smooth function of $z$). However, as can be seen in figure \ref{fig:nucleate}(b), after the ice lens has grown across the system, $dT_{\ell}/dz$ is discontinuous and equation (\ref{eqn:dTil}) shows that the volume flux is discontinuous with $U(z_n^+)\equiv U_1<U_0$ and $U(z_n^-) \equiv U_2>0$. This means that the flow upwards out of the lens is {\em{decreased}} while the flow upwards into the lens (from the reservoir) is {\em{increased}}. Therefore there is a net flow of water into the new ice lens that allows it to thicken. One implication of this is that there should be a surge in the heave rate of a column of soil (as measured by the volume flux of water being sucked up from the reservoir) each time that a new ice lens is formed, followed by a gradual decay until the point that the new lens initiates. This was observed experimentally by Penner \cite{penn86} and Akagawa et al. \cite{akag06}. It is interesting to note that the amplitude of the surge is controlled by the critical supercooling $T_mK_{Ic}/\rho L_m\sqrt{\pi\lambda R}$, and hence by the fracture toughness of the soil. Thus it may be possible to derive information about soil material parameters by measurement of the surge in heave rate. Additionally, no surge is predicted in a soil lacking cohesive strength.

This argument also demonstrates the link between geometrical supercooling and ice lens formation, namely the fact that the formation of a new ice lens should not be possible in a freezing soil if there is no geometrical supercooling of the pore water. If a new ice lens nucleates in the soil in which the pore liquid is not supercooled (e.g. figure \ref{fig:supercool}(b)), then an amended version of \ref{fig:nucleate}(b), taking account of a non-supercooled liquid, will show that $U_1>U_0$ and $U_2<U_0$ so there is a net flow of liquid {\em away} from the new lens. This means that the lens will reduce in thickness and disappear. Note that although a new ice lens cannot form without geometrical supercooling, it is still possible that ice will grow into the pores of the soil, forming a frozen fringe.

It is important to recall that we have assumed here that the soil is a rigid matrix so that this pressure adjustment is instantaneous, however for compressible materials the pressure adjustment will spread out diffusively from the position of the new ice lens over a finite time. The arguments presented above can be modified to take account of the transient pore pressure fields, and it will be seen that the qualitative conclusions are the same.

\section{Quantification of lens growth temperatures\label{section5}}

In order to demonstrate that this theory is quantitatively consistent with observations, in this section we estimate the temperature at which a new ice lens will form. We apply this estimate to two different materials for which experimental data is available: firstly, the silica microparticles used by Watanabe and coworkers \cite{wata00,wata02} and secondly, kaolinite clay, used in our own experiments on ice lens formation. In both of these cases, there exists experimental evidence that ice lenses can form without the presence of a frozen fringe \cite{wata00,brow90}. Thus pore ice is not expected to be present and so we use unfrozen values for the material fracture toughnesses.

\begin{figure}
\centering
\includegraphics[width=8cm]{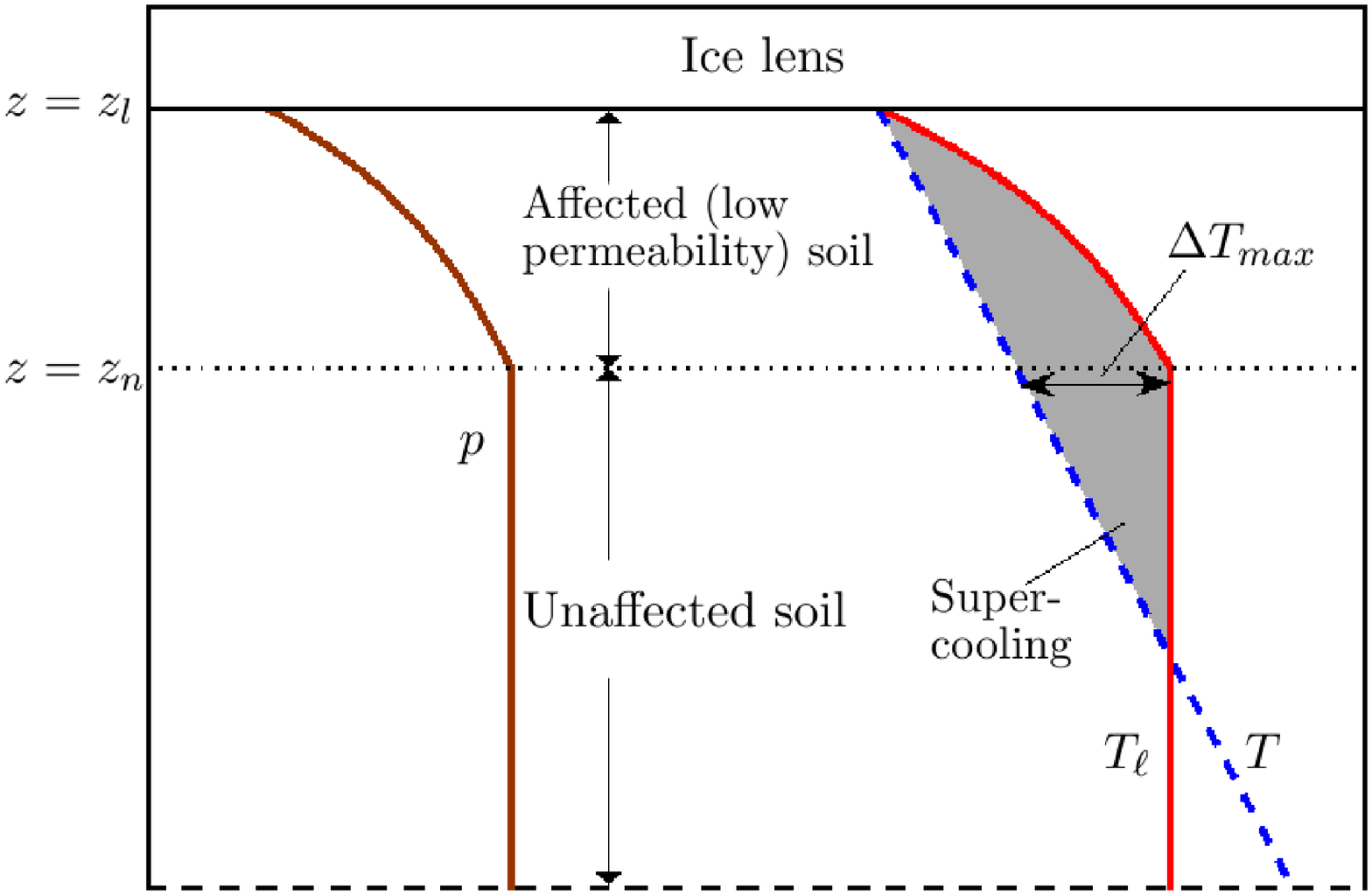}
\caption{(Color online) Schematic diagram showing the supercooling that occurs when the permeability of the soil is reduced in a region directly ahead of the ice lens. The affected region has a lower permeability due to the presence of pore ice, compression of the soil, or desaturation. The Darcy pressure is shown on the left hand side. On the right hand side, actual temperature (dashed curve) and ice-lens growth temperature (continuous curve) are shown. The maximum supercooling is approximately at the interface between unaffected and affected regions of soil. Compare with figure \ref{fig:supercool}.}
\label{fig:affectedarea}
\end{figure}

As described above, geometrical supercooling can be facilitated by reducing the permeability of the soil directly ahead of the warmest ice lens. We have suggested three possible mechanisms: (1) The appearance of a frozen fringe. (2) Having a highly compressible soil matrix. (3) Desaturation of the soil ahead of the ice lens. Each of these three mechanisms has the potential to reduce the soil permeability by at least an order of magnitude. For instance, Burt and Williams \cite{burt76} showed reductions in soil permeability of over four orders of magnitude as a frozen fringe formed. Peppin et al. \cite{pepp08} summarise evidence that the permeability of bentonite changes by over four orders of magnitude as it compresses. Finally, Peroni et al. \cite{pero03} report reductions in the permeability of kaolinite of over an order of magnitude upon desaturation. This means that the soil ahead of the warmest ice lens can be divided into two parts, as shown in figure \ref{fig:affectedarea}. Directly below the ice lens, there is an affected region with very low permeability, while further away from the lens, there is an unaffected region with high permeability. Provided that the low permeability region is not too thin, we can assume that the majority of the pressure drop between the reservoir $z=z_h$ and the ice lens $z=z_l$ occurs in the low permeability region, giving a pressure profile like that shown in figure \ref{fig:affectedarea} (the pressure is approximately constant in the unaffected soil region). This can be converted into $T_{\ell}$ using equation (\ref{eqn:Til}), and plotted against $T$ to determine the supercooling. Figure \ref{fig:affectedarea} shows that the maximum supercooling is expected to occur at the boundary between the affected and unaffected region of soil. At this point, $p\approx p_h$ and so, making use of equation (\ref{eqn:Til}) the geometrical supercooling is
\begin{equation}
T_{\ell}-T=T_m\left(1-\frac{P_o-p_h}{\rho L_m}\right)-T.
\end{equation}
In the experimental systems described above, the free surfaces are open to the air so that $P_o=p_h=P_a$, where $P_a$ is atmospheric pressure. Thus
\begin{equation}
 T_{\ell}-T \approx T_m-T,
\end{equation}
so the ice-lens nucleation criterion (\ref{eqn:supercool}) becomes
\begin{equation}
 T_m-T=\frac{T_m K_{Ic}}{\rho L_m \sqrt{\pi \lambda R}}=\frac{T_m \sigma_t}{\rho L_m}.
\end{equation}
Using this equation with figures for typical flaw sizes and material fracture toughness, we can estimate the undercooling $T_m-T$ at which a new lens will appear. It should be noted that this will serve as a lower bound for the undercooling, as with a less extreme pore pressure distribution than that shown in figure \ref{fig:affectedarea}, the system will need to be at a colder temperature before sufficient geometrical supercooling is present to form a new ice lens.

\begin{table}
\caption{Data specific to the experiments of Peppin et al. \cite{pepp07b,pepp08} and Watanabe \cite{wata02}. Approximate fracture toughnesses for silica microspheres and kaolinite clay are taken from Allain \& Limat \cite{alla95} and Ayad et al. \cite{ayad97} respectively. Particle sizes for silica microspheres and kaolinite clay are taken from \cite{wata02} and \cite{chin94} respectively.}
\label{table:tov}
\begin{ruledtabular}
\begin{tabular}{lrr}
Constant & Value & Units \\
\hline
$T_m$ & 273 & K \\
$\rho$ & $1 \times 10^3$ & $\mathrm{kg}\,\mathrm{m}^{-3}$ \\
$L_m$ & $3.3\times 10^5$ & $\rm{J}\,\rm{kg}^{-1}$\\
$K_{Ic}$ (Silica microspheres) & $1\times 10^2$ & Pa\,$\rm{m}^{1/2}$ \\
$K_{Ic}$ (Kaolinite) & $1.5\times 10^3$ & Pa\,$\rm{m}^{1/2}$ \\
$\lambda$ & 10 & - \\
$R$ (Silica microspheres) & 1.1$\times 10^{-6}$ & m \\
$R$ (Kaolinite) & 1$\times 10^{-6}$ & m \\
\end{tabular}
\end{ruledtabular}
\end{table}

If we substitute values into this equation from table \ref{table:tov} we find that in the silica microsphere system of Watanabe \cite{wata02}, new ice lenses are expected to form at an undercooling of $T_m-T\approx 0.014^\circ$C. This figure is within measurement precision of the minimum undercooling he recorded of $\sim0.02^\circ$C (that is, the warmest temperature at which ice lenses were observed). For kaolinite clay we find that new ice lenses are expected to form at an undercooling of $T_m-T\approx 0.22^\circ$C. Given the uncertainties in $K_{Ic}$, $\lambda$ and $R$ and the fact that our estimate represents a lower bound for the undercooling, the theory is in reasonable agreement with experiments, in which lenses were observed to form at temperatures of approximately $-0.3^\circ$C.

We are currently unaware of further data regarding ice-lens formation temperatures (for which the tensile properties of the material are available) against which we can test our theory, and therefore we must look to future experimental results to provide more rigorous validation. A simple verification would involve a systematic study of ice-lens formation temperatures in a variety of materials with different tensile strengths. At a more complex level, particularly useful experiments would be those that were able to tune the tensile strength of a soil while leaving other parameters, such as permeability, constant. As an example this might be achieved by using a material consisting of colloids with tunable interactions, or by altering the salt concentration in typical soils (while being aware of other potential changes that salt will make to the system, e.g. \cite{wors00}). Finally, it would obviously be of great benefit to be able to experimentally view the ice-lens formation process on the microscopic level, such as is routinely done for such systems in equilibrium \cite{dash06}, and we hope that methods of achieving this will become available in the near future.

\subsection{The presence of ice ahead of the warmest ice lens \label{section:51}}

A key part of our theory is the presence of ice in at least one large flaw ahead of the warmest existing ice lens, but there still remains the question of how ice can nucleate in suitable flaws. When there is a frozen fringe present, then ice can grow down through the soil pores to fill a flaw. However when there is no frozen fringe, there is no pore ice, and an alternative mechanism is required. In this case, there are two ways in which ice can grow into a flaw: homogeneous nucleation (spontaneous ice growth), and heterogeneous nucleation (growth from a pre-existing source of ice or from the wall of the flaw). We can rule out homogeneous nucleation in typical clays, as the temperatures at which ice lenses form (typically $\sim 0$ to $-0.5^\circ$C, e.g. \cite{penn86}) are much warmer than the temperatures at which ice spontaneously nucleates in clays ($-6^\circ$C or colder \cite{kozl09}). Therefore it seems likely that ice must be nucleated heterogeneously, and we suggest here two manners in which this may occur. Firstly, if the soil particles are good nucleators of ice (such as kaolinite), then ice can be nucleated from the sides of the flaw. In this case it is likely that ice will first nucleate in the largest flaws in the soil, for the same reasons that large droplets of water containing heterogeneous particles will solidify earlier than small droplets (e.g. \cite{lang58}). This will result in the situation shown in figure \ref{fig:flaw}.

\begin{figure}
\centering
\includegraphics[width=8cm]{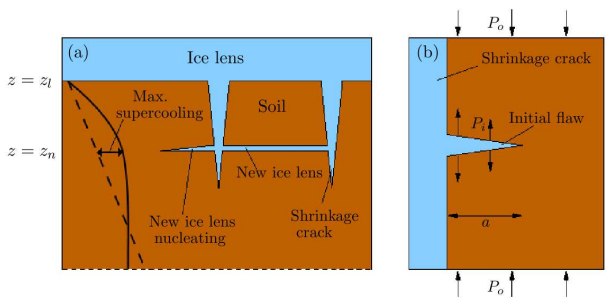}
\caption{(Color online) Schematic diagram of ice-lens formation in a compressible soil without a frozen fringe. Initially shrinkage cracks form and propagate downwards from the ice lens into the soil. New ice lenses can then nucleate from the shrinkage cracks in regions where sufficient geometrical supercooling exists. The solid and dashed lines show the ice-lens growth temperature and actual temperature respectively, with the new ice lens forming at the position of maximum supercooling, as in figure \ref{fig:supercool}.}
\label{fig:shrinkcrack}
\end{figure}

The second method by which we suggest that heterogeneous nucleation can occur is from the side of ice-filled shrinkage cracks that form ahead of a growing ice lens, as shown schematically in figure \ref{fig:shrinkcrack}(a). These shrinkage cracks are caused by suction in the soil adjacent to the growing ice lens, which puts the soil into tension, causing it to fracture in a manner analogous to the formation of mud cracks in drying soils \cite{cham79}. Similar to the case in drying soils, the shrinkage cracks in freezing soils generally take the form of a polygonal (in plan view) array of cracks propagating downwards through the soil, and these are commonly observed in tandem with ice-lens formation (e.g. \cite{cham79,pepp07b,aren08}). These shrinkage cracks provide a ready supply of ice ahead of the warmest ice lens; if a suitably large flaw exists in the soil on the side of a shrinkage crack, ice can grow into the flaw to form an edge crack, as shown schematically in figure \ref{fig:shrinkcrack}(b). A new ice lens can then nucleate through extension of the edge crack by the mechanism described previously. Interestingly, we have seen evidence for this behavior in our directional-solidification experiments on the freezing of kaolinite, an example of which is shown in figure \ref{fig:wata}.

When a new ice lens is nucleated from an edge crack, we need to make a small modification to account for the change of geometry from a center crack as shown in figure \ref{fig:flaw}, to an edge crack as shown in figure \ref{fig:shrinkcrack}(b). The deformation in the vicinity of a crack that breaks a free surface (an edge crack) is constrained less than that near a fully-embedded, center crack. Thus the crack opening for an edge crack is greater than that for a center crack for a given applied stress, and as $K$ scales with the opening, a higher stress intensity factor results. A detailed analysis of this problem (e.g. \cite{ande05}) then shows that stress intensity factor at the crack tip (\ref{eqn:sif}) is modified to become
\begin{equation}
K_I=1.12(P_i-P_o)\sqrt{\pi a}.
\end{equation}
Since $K$ for the edge crack is larger than that for a center crack, the critical condition for crack advance is reached at a lower value of internal pressure, which requires a lower degree of supercooling. This is seen in the modified supercooling condition, which becomes
\begin{equation}
T_{\ell}-T=\frac{T_m K_{Ic}}{1.12\rho L_m \sqrt{\pi \lambda R}}.
\end{equation}
Thus the critical supercooling for the edge crack is reduced from the critical supercooling calculated earlier (\ref{eqn:supercool}) by a factor of 0.89. This means that there will only be a small quantitative difference between the two cases.

In the case of nucleation off the side of a shrinkage crack, we also need to consider how the presence of shrinkage cracks affects $T_{\ell}$ through their effect on the Darcy pressure $p$. Shrinkage cracks appear in order to relieve the tensile stress that has built up in the soil, and then propagate steadily in front of the warmest ice lens. However, unlike ice lenses, which continue to thicken after they have nucleated, the shrinkage cracks remain the same thickness (e.g. \cite{aren08}). This means that there is no liquid being transported to, or away from the sides of the shrinkage cracks, and so the horizontal pressure gradient can be taken as zero. Thus the pressure is approximately one-dimensional, and should not be significantly affected by the presence of shrinkage cracks.

\section{Conclusions}

In this paper, we present a new, physically-intuitive theory of ice-lens nucleation. Motivated by experimental observations of crack-like behavior during the growth of ice lenses, we have determined the conditions under which an ice-filled flaw will split open. Ice in the flaw exerts an outwards pressure on the walls of the flaw, and when this pressure exceeds a critical value, we have shown that the flaw will extend horizontally across the soil to form a new ice lens.

We have introduced a new quantity, which we term the geometrical supercooling, that is intimately linked to the ice-lens growth process. We have shown that at any point in a soil, there is a certain temperature $T_{\ell}$ (referred to as the ice-lens growth temperature) below which energy will be released if a new ice lens were to form. However the cohesive nature of the soil opposes its tearing to form a new lens, and so the soil can exist at temperatures below $T_{\ell}$ without new lenses spontaneously forming. The geometrical supercooling, $T_{\ell}-T$, represents the cooling of the soil below this temperature $T_{\ell}$. We have shown that when the supercooling exceeds a critical amount, the ice pressure acting on the walls of an ice-filled flaw becomes sufficiently large to extend it, driving rapid horizontal growth to form a new ice lens. Furthermore, we have shown that the existence of geometrical supercooling is necessary for the formation of ice lenses. Our analysis suggests that if $T_{\ell}-T<0$, although pore ice may form, a new lens cannot appear regardless of the ice-lens nucleation mechanism. This is because if a new lens were to form, pressure gradients would immediately be set up so that it would melt away.

The geometrical supercooling necessary for the growth of ice lenses arises when the permeability of the soil directly ahead of the newest ice lens is low in comparison to the permeability of the warmer soil away from the freezing front. We have suggested three main scenarios by which this may occur. Firstly, as is commonly assumed in frost heave theories, a frozen fringe may penetrate the pores of the soil directly ahead of the warmest ice lens. Then the ice in the pores substantially reduces the soil permeability to water flow. Secondly, if the soil is highly compressible (such as a swelling clay like bentonite) then particles will be compressed together ahead of the ice lens, reducing the pore space, and thus the permeability. Finally, desaturation of the soil can occur ahead of the growing ice lens. This may be either due to bubble formation caused by air rejection from the growing ice lens, or due to cavitation caused by the large suctions that develop in the liquid. The desaturated region will be much less permeable than the saturated soil further from the lens. The highlighting of these three scenarios by no means precludes the existence of other mechanisms by which geometrical supercooling will occur. However at least one of these three is expected to occur in most freezing experiments, which indicates that geometrical supercooling is relatively common in freezing soils.

We have demonstrated that our theory is quantitatively in line with the available experimental data by calculating the warmest temperature at which new ice lenses are predicted to appear in two separate systems, the freezing of silica beads and of kaolinite clay. We hope that focused experimental studies of the ice-lens growth process will provide data that can further help verify our theory, and to this end we have suggested a number of experiments which would be helpful in analyzing the ice-growth process. Our theory is also in good qualitative agreement with observations. The theory explains the crack-like appearance of ice lenses as they grow, and also explains the experimentally-observed surge in the heave rate as each new lens forms (e.g. \cite{penn86}). Importantly, the theory explains the experimental finding that it is possible for discrete, periodic lenses to form even without a frozen fringe. We demonstrate that a particular mechanism by which this can occur is linked to the growth of ice-filled shrinkage cracks into the soil ahead of the warmest ice lens. After shrinkage crack growth, ice lenses can nucleate off the side of the shrinkage cracks to form the type of patterns shown schematically in figure \ref{fig:shrinkcrack} and observed in experiments on freezing clays (e.g. \cite{pepp07b, aren08, cham79}). Interestingly, in the case of ice lens growth without a frozen fringe, we have shown that the ice-lens formation can be predicted using only the unfrozen physical properties of the soil such as permeability and fracture toughness, which are readily experimentally-measurable.

In order to clearly demonstrate the physics at work we have restricted ourselves to a simple system with a linear temperature gradient. However the theory does not rely on this assumption, and can be readily generalized to nonlinear temperature distributions in the soil. Furthermore, the condition for ice-lens nucleation can be straightforwardly extended to higher dimensions, and thus the model should be useful in analyzing more involved situations in order to predict heave rates and ice-segregation. We hope that this work will be of interest to scientists across many disciplines, not only for research into the mesoscopic properties of freezing soil, but also due to its potential for up-scaling for the calculation of heave rates on macroscopic scales. For instance, we anticipate this to be useful in predicting the potential impacts of frost heave on engineering structures in cold climates, and also in exploring differential frost heave and the ability of freezing soils to create complex patterns such as stone circles and patterned ground. Finally, it should be noted that although we have focused on the growth of ice lenses in freezing soils, this theory is applicable to the freezing of dense colloidal suspensions and other porous media. Thus this work may also be of use in the analysis of ice segregation in porous media such as gels and rocks, and in colloidal systems such as are used in the development of biomaterials and other materials with specialized engineering properties \cite{devi07}.

\appendix
\section{The Application of the Clapeyron equation to freezing soils}

In a porous medium, the Darcy pressure is defined as follows: at the point where we wish to measure the pressure, we connect the pore fluid to a large reservoir of water and control the pressure of the reservoir until there is no flow between the pore fluid and the reservoir (i.e. they are in mechanical equilibrium). Then the Darcy pressure is given by the pressure in the reservoir, $p$ (e.g. \cite{pepp05}). There are several benefits of using the Darcy pressure over the actual liquid pressure that would be directly measured within the pores of the porous medium. Firstly, use of the Darcy pressure means that we do not have to worry about details of the geometry of the porous medium, or about the local effects of factors such as van der Waals and electrostatic interactions between the solid and liquid components in the pores. These factors may cause the {\em local} pore pressure to vary over sub-pore lengthscales, making this pressure hard to define on a macroscopic scale. However the Darcy pressure is well-defined macroscopically. Additionally, the flow through the porous medium is related to the gradient in the pressure by Darcy's law
\begin{equation}
 v=-\frac{k}{\mu}\nabla p.
\end{equation}
Secondly, although first derived for flow through saturated porous media, Darcy's law still holds if air bubbles form or ice nucleates in the pores of the porous medium, provided that the permeability $k$ is replaced by a new permeability that depends on the connectivity and amount of the water remaining in the porous medium. This is often used in the study of unsaturated soils (e.g. \cite{fred93})

Finally, by careful treatment of the thermodynamical problem the Darcy pressure within the soil can be related to the ice pressure in certain cases. If we consider thermodynamical equilibrium between ice and water at the same temperature, the pressure difference between the ice pressure and the liquid pressure is given by the modified Clapeyron equation
\begin{equation}
 P_i-p=\rho L \frac{(T_m-T)}{T_m},
\label{eqn:appa1}
\end{equation}
where we have ignored the small effects due to the density differences between ice and water (e.g. \cite{blac95}). It is important to emphasize that this equation is derived for equilibrium between {\em bulk} phases. This is because for small volumes of ice or water, the effects of surface forces acting on the volume are important, and these will affect the equation above. For instance, consider a pore filled with ice in a liquid-saturated porous medium. The edge of the ice consists of ice/premelted film/solid interfaces where the ice meets particles (e.g. \cite{remp04}), and ice/water interfaces where the ice meets pore throats between the particles. A first effect is due to van der Waals and electrostatic interactions between the ice and the particles/water. There will be an influence of these forces on the pressure in the ice over a lengthscale $\delta$, which is typically small in comparison with the size of a colloidal particle \cite{dash06}. A second effect is caused by the liquid and the particles surrounding the ice each transmitting different stresses to the pore ice. This results in a non-uniform pressure distribution in the ice near the pore walls. For a sufficiently large pore, this pressure distribution will average out away from the pore walls leaving a uniform pressure field $P_i$ in the middle of the ice, with the averaging occurring over a lengthscale comparable to the size of a soil particle $R$. Thus we speak of the bulk pressure $P_i$ in an ice pore if the size of the ice-filled pore is much bigger than $\delta$ and $R$. In effect, this means that ice in pores much larger than $R$ in size can be thought of as bulk ice.

Now consider an ice lens, such as that shown in figure \ref{fig:schem}, and the water in the pores directly adjacent to the lens. The lens is much larger than an individual particle size, and therefore can be treated as bulk ice. We connect up a reservoir to the pore fluid to measure the Darcy pressure and allow the system to come to local thermodynamic equilibrium. Then the bulk water in the reservoir is in equilibrium with the bulk water in the ice lens and so the conditions are satisfied for equation (\ref{eqn:appa1}) to hold. From mechanical equilibrium, we know that the ice-lens pressure is given by the overburden pressure. Thus
\begin{equation}
 P_o-p=\rho L \frac{(T_m-T)}{T_m}
\end{equation}
at the ice lens.

Now as a second scenario, consider an ice-filled pore in a porous medium. Typically the size of the pore will be around the same size as the particles, and so we cannot treat the ice within the pore as bulk ice. However, if we have a large flaw that is much bigger than the individual particle size, the ice inside the flaw can be considered as bulk (as mentioned in the text, there is evidence to suggest that these large flaws are present in most soils). Then equation (\ref{eqn:appa1}) holds, and if we know the local temperature and Darcy pressure of the liquid adjacent to the pore, we can calculate the ice pressure $P_i$ as with the case of the ice lens above. This is the basis of our calculation of the force acting to open a new ice lens.

As a final note, we also need to justify our assumption that the bulk pressure in the ice is isotropic within the flaw (i.e., that the ice is not under differential stress). For a solid under differential stress, the melting point of a face of the solid varies depending on the stress at the face \cite{seke04}. A consequence of this is that if such a solid is placed into a liquid, it cannot be in thermodynamic equilibrium with the liquid. If one side of the solid is in equilibrium with the liquid, then another side under a different stress will have a different melting temperature. This means that the differentially stressed solid is unstable with respect to the isotropically stressed solid and so any liquid and solid in equilibrium must both be in an isotropic stress state (i.e., the stress can be characterized by a pressure) \cite{seke04}. Because for slow freezing rates typical of frost-heaving soils pore ice is in equilibrium with surrounding pore liquid which is characterized by the isotropic stress (Darcy pressure) $p$, the pore ice must be under isotropic stress, unless external stresses are later applied to the soil column.

\section{The relationship between geometrical supercooling and constitutional supercooling}

As some readers may be familiar with the concept of constitutional supercooling in the freezing of suspensions and alloys, it is worth demonstrating the links between constitutional and geometrical supercooling. Constitutional supercooling is a phenomenon that has been well-studied for its role in causing morphological instabilities in the freezing of alloys and suspensions of colloidal particles \cite{wors00,pepp07b}. In a suspension of particles, the freezing temperature $T_f$ of the suspension is given by
\begin{equation}
 \rho L_m\frac{(T_m-T_f)}{T_m}=\Pi(\phi),
\label{eqn:osm}
\end{equation}
where $\Pi$ is its osmotic pressure, which depends on the volume fraction of solid particles $\phi$ \cite{pepp07b}. This equation implies that the freezing temperature is a function of particle volume fraction only: $T_f=T_f(\phi)$. If we know the temperature and concentration of a particle suspension, then the temperature $T$ can then be compared to the freezing temperature, and if $T<T_f$, the suspension is said to be constitutionally supercooled. The name `constitutional supercooling' therefore can be seen to come from the fact that the amount of supercooling is dependent on the constitution (i.e. concentration) of the particle suspension (or solution in the case of freezing alloys). A typical instance where this is known to occur is during the rapid freezing of colloidal suspensions. Colloidal particles are rejected from the advancing ice interface and build up in a boundary layer ahead of the ice, and if freezing is rapid enough constitutional supercooling will occur. This manifests itself as a morphological instability of the ice/suspension interface (e.g. \cite{pepp08}).

The analogy with geometrical supercooling comes from the use of the definition that $\Pi=P-p$ for a colloidal suspension, where $P$ is the overall, bulk pressure of the suspension and $p$ is the Darcy pressure. If we insert this expression into equation (\ref{eqn:osm}), and recognize that the bulk pressure of a suspension is the equivalent of the overburden pressure $P_o$ in a soil, we obtain
\begin{equation}
 T_f=T_m\left(1-\frac{P_o-p}{\rho L_m}\right).
\end{equation}
By comparison with equation (\ref{eqn:Til}), it can then be seen that $T_{\ell}$ is the generalization of $T_f$ to porous media. Thus geometrical supercooling is the extension of constitutional supercooling to porous media.

Although there is the strong analogy between the two supercoolings, geometrical supercooling is a much more general concept than constitutional supercooling when considering colloidal systems. Constitutional supercooling is restricted to the case where a freezing suspension has a well defined osmotic pressure as a function of $\phi$, so that equation (\ref{eqn:osm}) can be used. This means that it only applies to compressible suspensions of colloidal particles, and cannot be extended to incompressible, packed soils. Furthermore constitutional supercooling only assumes the presence of one `impurity' in the solute (in this case colloidal particles). On the other hand, geometrical supercooling is applicable to compressible and incompressible soils, and we have shown that it takes into account the presence of further impurities in the system such as air or pore ice. That pore ice can exist at the same time as supercooling is the main reason that we have chosen to call this phenomenon `geometrical supercooling'. With constitutional supercooling, the suspension is supercooled relative to the temperature at which ice can first appear, and the nucleation of ice will immediately cause ice to grow rapidly into the supercooled region, thereby removing the supercooling \cite{pepp07b}. However, as we have discussed in this paper, in a porous medium with a cohesive strength, pore ice can exist stably while the system is supercooled. Thus it is important to note that the system is not supercooled relative to the temperature at which ice can first grow, as microscopic, pore ice can already be present. Instead the system is supercooled relative to the macroscopic ice configuration.

\begin{acknowledgments}
This publication was based on work supported in part by Award No
KUK-C1-013-04, made by King Abdullah University of Science and
Technology (KAUST).  JSW thanks OCCAM for support as a Visiting 
Fellow, the John Simon Guggenheim Foundation and Yale University. 
\end{acknowledgments}

%


\begin{thebibliography}{10}%
\makeatletter
\providecommand \@ifxundefined [1]{%
 \ifx #1\undefined \expandafter \@firstoftwo
 \else \expandafter \@secondoftwo
\fi
}%
\providecommand \@ifnum [1]{%
 \ifnum #1\expandafter \@firstoftwo
 \else \expandafter \@secondoftwo
\fi
}%
\providecommand \enquote [1]{``#1''}%
\providecommand \bibnamefont  [1]{#1}%
\providecommand \bibfnamefont [1]{#1}%
\providecommand \citenamefont [1]{#1}%
\providecommand\href[0]{\@sanitize\@href}%
\providecommand\@href[1]{\endgroup\@@startlink{#1}\endgroup\@@href}%
\providecommand\@@href[1]{#1\@@endlink}%
\providecommand \@sanitize [0]{\begingroup\catcode`\&12\catcode`\#12\relax}%
\@ifxundefined \pdfoutput {\@firstoftwo}{%
 \@ifnum{\z@=\pdfoutput}{\@firstoftwo}{\@secondoftwo}%
}{%
 \providecommand\@@startlink[1]{\leavevmode}%
 \providecommand\@@endlink[0]{}%
}{%
 \providecommand\@@startlink[1]{%
  \leavevmode
  \pdfstartlink
   attr{/Border[0 0 1 ]/H/I/C[0 1 1]}%
   user{/Subtype/Link/A<</Type/Action/S/URI/URI(#1)>>}%
  \relax
 }%
 \providecommand\@@endlink[0]{\pdfendlink}%
}%
\providecommand \url  [0]{\begingroup\@sanitize \@url }%
\providecommand \@url [1]{\endgroup\@href {#1}{\urlprefix}}%
\providecommand \urlprefix [0]{URL }%
\providecommand \Eprint[0]{\href }%
\@ifxundefined \urlstyle {%
  \providecommand \doi [1]{doi:\discretionary{}{}{}#1}%
}{%
  \providecommand \doi [0]{doi:\discretionary{}{}{}\begingroup
  \urlstyle{rm}\Url }%
}%
\providecommand \doibase [0]{http://dx.doi.org/}%
\providecommand \Doi[1]{\href{\doibase#1}}%
\providecommand \bibAnnote [3]{%
  \BibitemShut{#1}%
  \begin{quotation}\noindent
    \textsc{Key:}\ #2\\\textsc{Annotation:}\ #3%
  \end{quotation}%
}%
\providecommand \bibAnnoteFile [2]{%
  \IfFileExists{#2}{\bibAnnote {#1} {#2} {\input{#2}}}{}%
}%
\providecommand \typeout [0]{\immediate \write \m@ne }%
\providecommand \selectlanguage [0]{\@gobble}%
\providecommand \bibinfo [0]{\@secondoftwo}%
\providecommand \bibfield [0]{\@secondoftwo}%
\providecommand \translation [1]{[#1]}%
\providecommand \BibitemOpen[0]{}%
\providecommand \bibitemStop [0]{}%
\providecommand \bibitemNoStop [0]{.\EOS\space}%
\providecommand \EOS [0]{\spacefactor3000\relax}%
\providecommand \BibitemShut [1]{\csname bibitem#1\endcsname}%
\bibitem{will91}%
  \BibitemOpen
  \bibfield{author}{%
  \bibinfo {author} {\bibfnamefont{P.~J.}\ \bibnamefont{Williams}}\ and\
  \bibinfo {author} {\bibfnamefont{M.~W.}\ \bibnamefont{Smith}},\ }%
  \emph{\bibinfo {title} {The Frozen Earth: Fundamentals of Geocryology}}\
  (\bibinfo {publisher} {Cambridge University Press},\ \bibinfo {address}
  {Cambridge},\ \bibinfo {year} {1991})%
  \bibAnnoteFile{NoStop}{will91}%
\bibitem{dash06}%
  \BibitemOpen
  \bibfield{author}{%
  \bibinfo {author} {\bibfnamefont{J.~G.}\ \bibnamefont{Dash}}, \bibinfo
  {author} {\bibfnamefont{A.~W.}\ \bibnamefont{Rempel}},\ and\ \bibinfo
  {author} {\bibfnamefont{J.~S.}\ \bibnamefont{Wettlaufer}},\ }%
  \bibfield{journal}{%
  \bibinfo {journal} {Rev. Mod. Phys.}\ }%
  \textbf{\bibinfo {volume} {78}},\ \bibinfo {pages} {695} (\bibinfo {year}
  {2006})%
  \bibAnnoteFile{NoStop}{dash06}%
\bibitem{kess03}%
  \BibitemOpen
  \bibfield{author}{%
  \bibinfo {author} {\bibfnamefont{M.~A.}\ \bibnamefont{Kessler}}\ and\
  \bibinfo {author} {\bibfnamefont{B.~T.}\ \bibnamefont{Werner}},\ }%
  \bibfield{journal}{%
  \bibinfo {journal} {Science}\ }%
  \textbf{\bibinfo {volume} {299}},\ \bibinfo {pages} {380} (\bibinfo {year}
  {2003})%
  \bibAnnoteFile{NoStop}{kess03}%
\bibitem{thom01}%
  \BibitemOpen
  \bibfield{author}{%
  \bibinfo {author} {\bibfnamefont{B.~J.}\ \bibnamefont{Thomson}}\ and\
  \bibinfo {author} {\bibfnamefont{J.~W.}\ \bibnamefont{Head}},\ }%
  \bibfield{journal}{%
  \bibinfo {journal} {J. Geophys. Res.}\ }%
  \textbf{\bibinfo {volume} {106}},\ \bibinfo {pages} {23209} (\bibinfo {year}
  {2001})%
  \bibAnnoteFile{NoStop}{thom01}%
\bibitem{dimi99}%
  \BibitemOpen
  \bibfield{author}{%
  \bibinfo {author} {\bibfnamefont{A.}~\bibnamefont{DiMillio}},\ }%
  \emph{\bibinfo {title} {A quarter century of geotechnical research}},\
  \bibinfo {type} {Tech. Rep.}\ (\bibinfo {institution} {Federal Highway
  Administration, US Department of Transportation},\ \bibinfo {year} {1999})%
  \bibAnnoteFile{NoStop}{dimi99}%
\bibitem{tabe29}%
  \BibitemOpen
  \bibfield{author}{%
  \bibinfo {author} {\bibfnamefont{S.}~\bibnamefont{Taber}},\ }%
  \bibfield{journal}{%
  \bibinfo {journal} {J. Geol.}\ }%
  \textbf{\bibinfo {volume} {37}},\ \bibinfo {pages} {428} (\bibinfo {year}
  {1929})%
  \bibAnnoteFile{NoStop}{tabe29}%
\bibitem{wors99}%
  \BibitemOpen
  \bibfield{author}{%
  \bibinfo {author} {\bibfnamefont{M.~G.}\ \bibnamefont{Worster}}\ and\
  \bibinfo {author} {\bibfnamefont{J.~S.}\ \bibnamefont{Wettlaufer}},\ }%
  \enquote{\bibinfo {title} {Fluid dynamics at interfaces},}\ \ (\bibinfo
  {publisher} {CUP},\ \bibinfo {year} {1999})\ Chap.\ \bibinfo {chapter} {The
  fluid mechanics of premelted liquid films}, pp.\ \bibinfo {pages} {339--351}%
  \bibAnnoteFile{NoStop}{wors99}%
\bibitem{remp04}%
  \BibitemOpen
  \bibfield{author}{%
  \bibinfo {author} {\bibfnamefont{A.~W.}\ \bibnamefont{Rempel}}, \bibinfo
  {author} {\bibfnamefont{J.~S.}\ \bibnamefont{Wettlaufer}},\ and\ \bibinfo
  {author} {\bibfnamefont{M.~G.}\ \bibnamefont{Worster}},\ }%
  \bibfield{journal}{%
  \bibinfo {journal} {J.\ Fluid Mech.}\ }%
  \textbf{\bibinfo {volume} {498}},\ \bibinfo {pages} {227} (\bibinfo {year}
  {2004})%
  \bibAnnoteFile{NoStop}{remp04}%
\bibitem{pepp07b}%
  \BibitemOpen
  \bibfield{author}{%
  \bibinfo {author} {\bibfnamefont{S.~S.~L.}\ \bibnamefont{Peppin}}, \bibinfo
  {author} {\bibfnamefont{M.~G.}\ \bibnamefont{Worster}},\ and\ \bibinfo
  {author} {\bibfnamefont{J.~S.}\ \bibnamefont{Wettlaufer}},\ }%
  \bibfield{journal}{%
  \bibinfo {journal} {Proc. Roy. Soc. (London) A}\ }%
  \textbf{\bibinfo {volume} {463}},\ \bibinfo {pages} {723} (\bibinfo {year}
  {2007})%
  \bibAnnoteFile{NoStop}{pepp07b}%
\bibitem{pepp08}%
  \BibitemOpen
  \bibfield{author}{%
  \bibinfo {author} {\bibfnamefont{S.~S.~L.}\ \bibnamefont{Peppin}}, \bibinfo
  {author} {\bibfnamefont{J.~S.}\ \bibnamefont{Wettlaufer}},\ and\ \bibinfo
  {author} {\bibfnamefont{M.~G.}\ \bibnamefont{Worster}},\ }%
  \bibfield{journal}{%
  \bibinfo {journal} {Phys. Rev. Lett.}\ }%
  \textbf{\bibinfo {volume} {100}},\ \bibinfo {pages} {238301} (\bibinfo {year}
  {2008})%
  \bibAnnoteFile{NoStop}{pepp08}%
\bibitem{tabe30}%
  \BibitemOpen
  \bibfield{author}{%
  \bibinfo {author} {\bibfnamefont{S.}~\bibnamefont{Taber}},\ }%
  \bibfield{journal}{%
  \bibinfo {journal} {J. Geol.}\ }%
  \textbf{\bibinfo {volume} {38}},\ \bibinfo {pages} {303} (\bibinfo {year}
  {1930})%
  \bibAnnoteFile{NoStop}{tabe30}%
\bibitem{onei85}%
  \BibitemOpen
  \bibfield{author}{%
  \bibinfo {author} {\bibfnamefont{K.}~\bibnamefont{O'Neill}}\ and\ \bibinfo
  {author} {\bibfnamefont{R.~D.}\ \bibnamefont{Miller}},\ }%
  \bibfield{journal}{%
  \bibinfo {journal} {Water Resour. Res.}\ }%
  \textbf{\bibinfo {volume} {21}},\ \bibinfo {pages} {281} (\bibinfo {year}
  {1985})%
  \bibAnnoteFile{NoStop}{onei85}%
\bibitem{gilp80}%
  \BibitemOpen
  \bibfield{author}{%
  \bibinfo {author} {\bibfnamefont{R.~R.}\ \bibnamefont{Gilpin}},\ }%
  \bibfield{journal}{%
  \bibinfo {journal} {Water Resour. Res.}\ }%
  \textbf{\bibinfo {volume} {16}},\ \bibinfo {pages} {918} (\bibinfo {year}
  {1980})%
  \bibAnnoteFile{NoStop}{gilp80}%
\bibitem{fowl89}%
  \BibitemOpen
  \bibfield{author}{%
  \bibinfo {author} {\bibfnamefont{A.~C.}\ \bibnamefont{Fowler}},\ }%
  \bibfield{journal}{%
  \bibinfo {journal} {SIAM J. Appl. Maths}\ }%
  \textbf{\bibinfo {volume} {49}},\ \bibinfo {pages} {991} (\bibinfo {year}
  {1989})%
  \bibAnnoteFile{NoStop}{fowl89}%
\bibitem{chri06}%
  \BibitemOpen
  \bibfield{author}{%
  \bibinfo {author} {\bibfnamefont{P.}~\bibnamefont{Christofferson}}, \bibinfo
  {author} {\bibfnamefont{S.}~\bibnamefont{Tulaczyk}}, \bibinfo {author}
  {\bibfnamefont{F.~D.}\ \bibnamefont{Carsey}},\ and\ \bibinfo {author}
  {\bibfnamefont{A.~E.}\ \bibnamefont{Behar}},\ }%
  \bibfield{journal}{%
  \bibinfo {journal} {J. Geophys. Res.}\ }%
  \textbf{\bibinfo {volume} {111}},\ \bibinfo {pages} {F01017} (\bibinfo {year}
  {2006})%
  \bibAnnoteFile{NoStop}{chri06}%
\bibitem{wata00}%
  \BibitemOpen
  \bibfield{author}{%
  \bibinfo {author} {\bibfnamefont{K.}~\bibnamefont{Watanabe}}\ and\ \bibinfo
  {author} {\bibfnamefont{M.}~\bibnamefont{Mizoguchi}},\ }%
  \bibfield{journal}{%
  \bibinfo {journal} {J. Cryst. Growth}\ }%
  \textbf{\bibinfo {volume} {213}},\ \bibinfo {pages} {135} (\bibinfo {year}
  {2000})%
  \bibAnnoteFile{NoStop}{wata00}%
\bibitem{brow90}%
  \BibitemOpen
  \bibfield{author}{%
  \bibinfo {author} {\bibfnamefont{S.~C.}\ \bibnamefont{Brown}}\ and\ \bibinfo
  {author} {\bibfnamefont{D.}~\bibnamefont{Payne}},\ }%
  \bibfield{journal}{%
  \bibinfo {journal} {J. Soil. Sci.}\ }%
  \textbf{\bibinfo {volume} {41}},\ \bibinfo {pages} {547} (\bibinfo {year}
  {1990})%
  \bibAnnoteFile{NoStop}{brow90}%
\bibitem{cham79}%
  \BibitemOpen
  \bibfield{author}{%
  \bibinfo {author} {\bibfnamefont{E.~J.}\ \bibnamefont{Chamberlain}}\ and\
  \bibinfo {author} {\bibfnamefont{A.~J.}\ \bibnamefont{Gow}},\ }%
  \bibfield{journal}{%
  \bibinfo {journal} {Eng. Geology}\ }%
  \textbf{\bibinfo {volume} {13}},\ \bibinfo {pages} {73} (\bibinfo {year}
  {1979})%
  \bibAnnoteFile{NoStop}{cham79}%
\bibitem{aren08}%
  \BibitemOpen
  \bibfield{author}{%
  \bibinfo {author} {\bibfnamefont{L.~U.}\ \bibnamefont{Arenson}}, \bibinfo
  {author} {\bibfnamefont{T.~F.}\ \bibnamefont{Azmatch}},\ and\ \bibinfo
  {author} {\bibfnamefont{D.~C.}\ \bibnamefont{Sego}},\ }%
  in\ \emph{\bibinfo {booktitle} {Proceedings of the Ninth International
  Conference on Permafrost, University of Alaska Fairbanks}}\ (\bibinfo {year}
  {2008})%
  \bibAnnoteFile{NoStop}{aren08}%
\bibitem{akag06}%
  \BibitemOpen
  \bibfield{author}{%
  \bibinfo {author} {\bibfnamefont{S.}~\bibnamefont{Akagawa}}, \bibinfo
  {author} {\bibfnamefont{M.}~\bibnamefont{Satoh}}, \bibinfo {author}
  {\bibfnamefont{S.}~\bibnamefont{Kanie}},\ and\ \bibinfo {author}
  {\bibfnamefont{T.}~\bibnamefont{Mikami}},\ }%
  in\ \emph{\bibinfo {booktitle} {Current Practices in Cold Regions
  Engineering, Proceedings of the 13th International Conference on Cold Regions
  Engineering July 23-26, 2006, Orono, Maine}}\ (\bibinfo {year} {2006})%
  \bibAnnoteFile{NoStop}{akag06}%
\bibitem{wald85}%
  \BibitemOpen
  \bibfield{author}{%
  \bibinfo {author} {\bibfnamefont{J.}~\bibnamefont{Walder}}\ and\ \bibinfo
  {author} {\bibfnamefont{B.}~\bibnamefont{Hallet}},\ }%
  \bibfield{journal}{%
  \bibinfo {journal} {Geol. Soc. Am. Bull.}\ }%
  \textbf{\bibinfo {volume} {96}},\ \bibinfo {pages} {336} (\bibinfo {year}
  {1985})%
  \bibAnnoteFile{NoStop}{wald85}%
\bibitem{pepp07}%
  \BibitemOpen
  \bibfield{author}{%
  \bibinfo {author} {\bibfnamefont{S.~S.~L.}\ \bibnamefont{Peppin}}, \bibinfo
  {author} {\bibfnamefont{P.}~\bibnamefont{Aussillous}}, \bibinfo {author}
  {\bibfnamefont{H.~E.}\ \bibnamefont{Huppert}},\ and\ \bibinfo {author}
  {\bibfnamefont{M.~G.}\ \bibnamefont{Worster}},\ }%
  \bibfield{journal}{%
  \bibinfo {journal} {J. Fluid Mech.}\ }%
  \textbf{\bibinfo {volume} {570}},\ \bibinfo {pages} {69} (\bibinfo {year}
  {2007})%
  \bibAnnoteFile{NoStop}{pepp07}%
\bibitem{davi01}%
  \BibitemOpen
  \bibfield{author}{%
  \bibinfo {author} {\bibfnamefont{S.~H.}\ \bibnamefont{Davis}},\ }%
  \emph{\bibinfo {title} {Theory of Solidification}}\ (\bibinfo {publisher}
  {CUP},\ \bibinfo {year} {2001})%
  \bibAnnoteFile{NoStop}{davi01}%
\bibitem{blac95}%
  \BibitemOpen
  \bibfield{author}{%
  \bibinfo {author} {\bibfnamefont{P.~B.}\ \bibnamefont{Black}},\ }%
  \emph{\bibinfo {title} {Applications of the Clapeyron Equation to Water and
  Ice in Porous Media}},\ \bibinfo {type} {Tech. Rep.}\ (\bibinfo {institution}
  {U.S. Army Cold Regions Research and Engineering Laboratory},\ \bibinfo
  {year} {1995})%
  \bibAnnoteFile{NoStop}{blac95}%
\bibitem{ande05}%
  \BibitemOpen
  \bibfield{author}{%
  \bibinfo {author} {\bibfnamefont{T.~L.}\ \bibnamefont{Anderson}},\ }%
  \emph{\bibinfo {title} {Fracture Mechanics: Fundamentals and Applications}}\
  (\bibinfo {publisher} {Taylor \& Francis},\ \bibinfo {year} {2005})%
  \bibAnnoteFile{NoStop}{ande05}%
\bibitem{diam70}%
  \BibitemOpen
  \bibfield{author}{%
  \bibinfo {author} {\bibfnamefont{S.}~\bibnamefont{Diamond}},\ }%
  \bibfield{journal}{%
  \bibinfo {journal} {Clays Clay Min.}\ }%
  \textbf{\bibinfo {volume} {18}},\ \bibinfo {pages} {7} (\bibinfo {year}
  {1970})%
  \bibAnnoteFile{NoStop}{diam70}%
\bibitem{grah83}%
  \BibitemOpen
  \bibfield{author}{%
  \bibinfo {author} {\bibfnamefont{J.}~\bibnamefont{Graham}}\ and\ \bibinfo
  {author} {\bibfnamefont{G.~T.}\ \bibnamefont{Houlsby}},\ }%
  \bibfield{journal}{%
  \bibinfo {journal} {Geotechnique}\ }%
  \textbf{\bibinfo {volume} {33}},\ \bibinfo {pages} {165} (\bibinfo {year}
  {1983})%
  \bibAnnoteFile{NoStop}{grah83}%
\bibitem{muir90}%
  \BibitemOpen
  \bibfield{author}{%
  \bibinfo {author} {\bibfnamefont{D.~M.}\ \bibnamefont{Wood}},\ }%
  \emph{\bibinfo {title} {Soil behaviour and critical state soil mechanics}}\
  (\bibinfo {publisher} {Cambridge University Press},\ \bibinfo {year} {1990})%
  \bibAnnoteFile{NoStop}{muir90}%
\bibitem{wors86}%
  \BibitemOpen
  \bibfield{author}{%
  \bibinfo {author} {\bibfnamefont{M.~G.}\ \bibnamefont{Worster}},\ }%
  \bibfield{journal}{%
  \bibinfo {journal} {J.~Fluid~Mech.}\ }%
  \textbf{\bibinfo {volume} {167}},\ \bibinfo {pages} {481} (\bibinfo {year}
  {1986})%
  \bibAnnoteFile{NoStop}{wors86}%
\bibitem{hans10}%
  \BibitemOpen
  \bibfield{author}{%
  \bibinfo {author} {\bibfnamefont{H.}~\bibnamefont{Hansen-Goos}}\ and\
  \bibinfo {author} {\bibfnamefont{J.~S.}\ \bibnamefont{Wettlaufer}},\ }%
  \bibfield{journal}{%
  \bibinfo {journal} {Phys. Rev. E}\ }%
  \textbf{\bibinfo {volume} {81}},\ \bibinfo {pages} {031604} (\bibinfo {year}
  {2010})%
  \bibAnnoteFile{NoStop}{hans10}%
\bibitem{pepp06}%
  \BibitemOpen
  \bibfield{author}{%
  \bibinfo {author} {\bibfnamefont{S.~S.~L.}\ \bibnamefont{Peppin}}, \bibinfo
  {author} {\bibfnamefont{J.~A.~W.}\ \bibnamefont{Elliott}},\ and\ \bibinfo
  {author} {\bibfnamefont{M.~G.}\ \bibnamefont{Worster}},\ }%
  \bibfield{journal}{%
  \bibinfo {journal} {J. Fluid Mech.}\ }%
  \textbf{\bibinfo {volume} {554}},\ \bibinfo {pages} {147} (\bibinfo {year}
  {2006})%
  \bibAnnoteFile{NoStop}{pepp06}%
\bibitem{or02}%
  \BibitemOpen
  \bibfield{author}{%
  \bibinfo {author} {\bibfnamefont{D.}~\bibnamefont{Or}}\ and\ \bibinfo
  {author} {\bibfnamefont{M.}~\bibnamefont{Tuller}},\ }%
  \bibfield{journal}{%
  \bibinfo {journal} {Water Resour. Res.}\ }%
  \textbf{\bibinfo {volume} {38}},\ \bibinfo {pages} {1061} (\bibinfo {year}
  {2002})%
  \bibAnnoteFile{NoStop}{or02}%
\bibitem{penn63}%
  \BibitemOpen
  \bibfield{author}{%
  \bibinfo {author} {\bibfnamefont{E.}~\bibnamefont{Penner}},\ }%
  in\ \emph{\bibinfo {booktitle} {Proceedings: permafrost international
  conference}}\ (\bibinfo {year} {1963})\ pp.\ \bibinfo {pages} {197--202}%
  \bibAnnoteFile{NoStop}{penn63}%
\bibitem{fred93}%
  \BibitemOpen
  \bibfield{author}{%
  \bibinfo {author} {\bibfnamefont{D.~G.}\ \bibnamefont{Fredlund}}\ and\
  \bibinfo {author} {\bibfnamefont{H.}~\bibnamefont{Rahardjo}},\ }%
  \emph{\bibinfo {title} {Soil mechanics for unsaturated soils}}\ (\bibinfo
  {publisher} {John Wiley \& Sons},\ \bibinfo {year} {1993})%
  \bibAnnoteFile{NoStop}{fred93}%
\bibitem{penn86}%
  \BibitemOpen
  \bibfield{author}{%
  \bibinfo {author} {\bibfnamefont{E.}~\bibnamefont{Penner}},\ }%
  \bibfield{journal}{%
  \bibinfo {journal} {Cold Reg. Sci. Tech.}\ }%
  \textbf{\bibinfo {volume} {13}},\ \bibinfo {pages} {91} (\bibinfo {year}
  {1986})%
  \bibAnnoteFile{NoStop}{penn86}%
\bibitem{wata02}%
  \BibitemOpen
  \bibfield{author}{%
  \bibinfo {author} {\bibfnamefont{K.}~\bibnamefont{Watanabe}},\ }%
  \bibfield{journal}{%
  \bibinfo {journal} {J. Cryst. Growth}\ }%
  \textbf{\bibinfo {volume} {237-239}},\ \bibinfo {pages} {2194} (\bibinfo
  {year} {2002})%
  \bibAnnoteFile{NoStop}{wata02}%
\bibitem{burt76}%
  \BibitemOpen
  \bibfield{author}{%
  \bibinfo {author} {\bibfnamefont{T.~P.}\ \bibnamefont{Burt}}\ and\ \bibinfo
  {author} {\bibfnamefont{P.~J.}\ \bibnamefont{Williams}},\ }%
  \bibfield{journal}{%
  \bibinfo {journal} {Earth Surf. Proc.}\ }%
  \textbf{\bibinfo {volume} {1}},\ \bibinfo {pages} {349} (\bibinfo {year}
  {1976})%
  \bibAnnoteFile{NoStop}{burt76}%
\bibitem{pero03}%
  \BibitemOpen
  \bibfield{author}{%
  \bibinfo {author} {\bibfnamefont{N.}~\bibnamefont{Peroni}}, \bibinfo {author}
  {\bibfnamefont{E.}~\bibnamefont{Fratalocchi}},\ and\ \bibinfo {author}
  {\bibfnamefont{A.}~\bibnamefont{Tarantino}},\ }%
  in\ \emph{\bibinfo {booktitle} {Unsaturated Soils: Experimental Studies.
  Proceedings of the international conference "From experimental evidence
  towards numerical modelling of unsaturated soils," Weimar, Germany.}}\
  (\bibinfo {year} {2003})%
  \bibAnnoteFile{NoStop}{pero03}%
\bibitem{alla95}%
  \BibitemOpen
  \bibfield{author}{%
  \bibinfo {author} {\bibfnamefont{C.}~\bibnamefont{Allain}}\ and\ \bibinfo
  {author} {\bibfnamefont{L.}~\bibnamefont{Limat}},\ }%
  \bibfield{journal}{%
  \bibinfo {journal} {Phys. Rev. Lett.}\ }%
  \textbf{\bibinfo {volume} {74}},\ \bibinfo {pages} {2981} (\bibinfo {year}
  {1995})%
  \bibAnnoteFile{NoStop}{alla95}%
\bibitem{ayad97}%
  \BibitemOpen
  \bibfield{author}{%
  \bibinfo {author} {\bibfnamefont{R.}~\bibnamefont{Ayad}}, \bibinfo {author}
  {\bibfnamefont{J.~M.}\ \bibnamefont{Konrad}},\ and\ \bibinfo {author}
  {\bibfnamefont{M.}~\bibnamefont{Soulie}},\ }%
  \bibfield{journal}{%
  \bibinfo {journal} {Can. Geotech. J.}\ }%
  \textbf{\bibinfo {volume} {34}},\ \bibinfo {pages} {943} (\bibinfo {year}
  {1997})%
  \bibAnnoteFile{NoStop}{ayad97}%
\bibitem{chin94}%
  \BibitemOpen
  \bibfield{author}{%
  \bibinfo {author} {\bibfnamefont{H.~W.}\ \bibnamefont{Ching}}, \bibinfo
  {author} {\bibfnamefont{T.~S.}\ \bibnamefont{Tanaka}},\ and\ \bibinfo
  {author} {\bibfnamefont{M.}~\bibnamefont{Elimelech}},\ }%
  \bibfield{journal}{%
  \bibinfo {journal} {Wat. Res.}\ }%
  \textbf{\bibinfo {volume} {28}},\ \bibinfo {pages} {559} (\bibinfo {year}
  {1994})%
  \bibAnnoteFile{NoStop}{chin94}%
\bibitem{kozl09}%
  \BibitemOpen
  \bibfield{author}{%
  \bibinfo {author} {\bibfnamefont{T.}~\bibnamefont{Kozlowski}},\ }%
  \bibfield{journal}{%
  \bibinfo {journal} {Cold Reg. Sci. Tech.}\ }%
  \textbf{\bibinfo {volume} {59}},\ \bibinfo {pages} {25} (\bibinfo {year}
  {2009})%
  \bibAnnoteFile{NoStop}{kozl09}%
\bibitem{lang58}%
  \BibitemOpen
  \bibfield{author}{%
  \bibinfo {author} {\bibfnamefont{E.~J.}\ \bibnamefont{Langham}}\ and\
  \bibinfo {author} {\bibfnamefont{B.~J.}\ \bibnamefont{Mason}},\ }%
  \bibfield{journal}{%
  \bibinfo {journal} {Proc. Roy. Soc. Lond. A}\ }%
  \textbf{\bibinfo {volume} {247}},\ \bibinfo {pages} {493} (\bibinfo {year}
  {1958})%
  \bibAnnoteFile{NoStop}{lang58}%
\bibitem{devi07}%
  \BibitemOpen
  \bibfield{author}{%
  \bibinfo {author} {\bibfnamefont{S.}~\bibnamefont{Deville}}, \bibinfo
  {author} {\bibfnamefont{S.}~\bibnamefont{Saiz}},\ and\ \bibinfo {author}
  {\bibfnamefont{A.~P.}\ \bibnamefont{Tomsia}},\ }%
  \bibfield{journal}{%
  \bibinfo {journal} {Acta Mater.}\ }%
  \textbf{\bibinfo {volume} {55}},\ \bibinfo {pages} {1965} (\bibinfo {year}
  {2007})%
  \bibAnnoteFile{NoStop}{devi07}%
\bibitem{pepp05}%
  \BibitemOpen
  \bibfield{author}{%
  \bibinfo {author} {\bibfnamefont{S.~S.~L.}\ \bibnamefont{Peppin}}, \bibinfo
  {author} {\bibfnamefont{J.~A.~W.}\ \bibnamefont{Elliott}},\ and\ \bibinfo
  {author} {\bibfnamefont{M.~G.}\ \bibnamefont{Worster}},\ }%
  \bibfield{journal}{%
  \bibinfo {journal} {Phys. Fluids}\ }%
  \textbf{\bibinfo {volume} {17}},\ \bibinfo {pages} {053301} (\bibinfo {year}
  {2005})%
  \bibAnnoteFile{NoStop}{pepp05}%
\bibitem{seke04}%
  \BibitemOpen
  \bibfield{author}{%
  \bibinfo {author} {\bibfnamefont{R.~F.}\ \bibnamefont{Sekerka}}\ and\
  \bibinfo {author} {\bibfnamefont{J.~W.}\ \bibnamefont{Cahn}},\ }%
  \bibfield{journal}{%
  \bibinfo {journal} {Acta Mater.}\ }%
  \textbf{\bibinfo {volume} {52}},\ \bibinfo {pages} {1663} (\bibinfo {year}
  {2004})%
  \bibAnnoteFile{NoStop}{seke04}%
\bibitem{wors00}%
  \BibitemOpen
  \bibfield{author}{%
  \bibinfo {author} {\bibfnamefont{M.~G.}\ \bibnamefont{Worster}},\ }%
  in\ \emph{\bibinfo {booktitle} {Perspectives in Fluid Dynamics}},\ \bibinfo
  {editor} {edited by\ \bibinfo {editor} {\bibfnamefont{G.~K.}\
  \bibnamefont{Batchelor}}, \bibinfo {editor} {\bibfnamefont{H.~K.}\
  \bibnamefont{Moffat}},\ and\ \bibinfo {editor} {\bibfnamefont{M.~G.}\
  \bibnamefont{Worster}}}\ (\bibinfo {publisher} {Cambridge University Press},\
  \bibinfo {year} {2000})%
  \bibAnnoteFile{NoStop}{wors00}%
\end{thebibliography}
\end{document}